%% file: ms_3.tex
\documentclass{elsart}

\def  \betabf  {{\mbox{\boldmath$\beta$}}}

\usepackage{natbib}
\usepackage{amssymb}
\usepackage{lscape}

\journal{Accident Analysis and Prevention}\date{}

\begin{document}

\begin{frontmatter}


\title{Markov switching multinomial logit model:
an application to accident injury severities}

\author{Nataliya V. Malyshkina\corauthref{correspond_author}},
\corauth[correspond_author]{Corresponding author.}
\ead{nmalyshk@purdue.edu}
\author{Fred L. Mannering}
\ead{flm@ecn.purdue.edu}
\address{School of Civil Engineering, 550 Stadium Mall Drive,
Purdue University, West Lafayette, IN 47907, United States}

\begin{abstract}
In this study, two-state Markov switching multinomial logit models
are proposed for statistical modeling of accident injury severities.
These models assume Markov switching in time between two unobserved
states of roadway safety. The states are distinct,
in the sense that in different states accident severity outcomes are
generated by separate multinomial logit processes. To demonstrate the
applicability of the approach presented herein, two-state Markov
switching multinomial logit models are estimated for severity outcomes
of accidents occurring on Indiana roads over a four-year time
interval. Bayesian inference methods and Markov Chain Monte Carlo
(MCMC) simulations are used for model estimation. The estimated Markov
switching models result in a superior statistical fit relative
to the standard (single-state) multinomial logit models. It is
found that the more frequent state of roadway safety is
correlated with better weather conditions. The less frequent
state is found to be correlated with adverse weather conditions.
\end{abstract}


\end{frontmatter}

\section{Introduction}
\label{INTRO}

Vehicle accidents result in property damage, injuries and loss of
people lives. Thus, research efforts in predicting accident severity
are clearly very important. In the past there has been a large
number of studies that focused on modeling accident severity
outcomes. Common modeling approaches of accident severity include
multinomial logit models, nested logit models, mixed logit models
and ordered probit models \citep[][]{OC_96,SM_96,SMB_96,DKC_98,
CM_99,CM_01,K_01,KPSH_02,KK_02,LM_02,A_03,KK_03,UM_04,YS_04,
KNSM_05,EB_07,SM_07,MSM_08}.
All these models involve nonlinear regression of the observed
accident injury severity outcomes on various accident characteristics
and related factors (such as roadway and driver characteristics,
environmental factors, etc).

In our earlier paper, \citet[][]{MMT_08}, which we will refer to as
Paper~I, we presented two-state Markov switching count data models
of accident frequencies. In this study, which is a continuation of
our work on Markov switching models, we present two-state Markov
switching multinomial logit models for predicting accident severity
outcomes. These models assume that there are two unobserved states
of roadway safety, roadway entities (roadway segments)
can switch between these states over time, and the
switching process is Markovian. The two states intend to account
for possible heterogeneity effects in roadway safety, which may be
caused by various unpredictable, unidentified, unobservable
risk factors that influence roadway safety. Because the risk factors
can interact and change, roadway entities can switch between the two
states over time. Two-state Markov switching multinomial logit models
assume separate multinomial logit processes for accident severity
data generation in the two states and, therefore, allow a researcher
to study the heterogeneity effects in roadway safety.

\section{Model specification}
\label{MOD_SPECIF}
Markov switching models are parametric and can be fully specified
by a likelihood function $f({\bf Y}|{\bf\Theta},{\cal M})$,
which  is the conditional probability
distribution of the vector of all observations ${\bf Y}$, given
the vector of all parameters ${\bf\Theta}$ of model ${\cal M}$.
First, let us consider ${\bf Y}$. Let $N_t$ be the number of
accidents observed during time period $t$, where $t=1,2,\ldots,T$
and $T$ is the total number of time periods. Let there be $I$
discrete outcomes observed for accident severity (for example,
$I=3$ and these outcomes are fatality, injury and property damage
only). Let us introduce accident severity outcome dummies
$\delta^{(i)}_{t,n}$ that are equal to unity if the $i^{\rm th}$
severity outcome is observed in the $n^{\rm th}$ accident that occurs
during time period $t$, and to zero otherwise. Here $i=1,2,\ldots,I$,
$n=1,2,\ldots,N_t$ and $t=1,2,\ldots,T$. Then, our observations are
the accident severity outcomes, and the vector of all observations
${\bf Y}=\{\delta^{(i)}_{t,n}\}$ includes all outcomes observed in
all accidents that occur during all time periods.
Second, let us consider model specification variable ${\cal M}$.
It is ${\cal M}=\{M,{\bf X}_{t,n}\}$ and includes
the model's name $M$ (for example, $M=\mbox{``multinomial logit''}$)
and the vector ${\bf X}_{t,n}$ of all accident characteristic
variables (weather and environment conditions, vehicle and driver
characteristics, roadway and pavement properties, and so on).

To define the likelihood function, we first introduce an
unobserved (latent) state variable $s_t$, which determines
the state of all roadway entities during time period $t$.
At each $t$, the state variable $s_t$ can assume only two values:
$s_t=0$ corresponds to one state and $s_t=1$ corresponds to the
other state ($t=1,2,\ldots,T$). The state variable $s_t$ is assumed
to follow a stationary two-state Markov chain process in
time,\footnote{\label{MARKOV} Markov property means that the probability
distribution of $s_{t+1}$ depends only on the value $s_{t}$ at time $t$,
but not on the previous history $s_{t-1},s_{t-2},\ldots$. Stationarity
of $\{s_t\}$ is in the statistical sense.}
which can be specified by time-independent transition probabilities as
\begin{eqnarray}
P(s_{t+1}=1|s_t=0)=p_{0\rightarrow1},
\quad P(s_{t+1}=0|s_t=1)=p_{1\rightarrow0}.
\label{EQ_P}
\end{eqnarray}
Here, for example, $P(s_{t+1}=1|s_t=0)$ is the conditional probability
of $s_{t+1}=1$ at time $t+1$, given that $s_t=0$ at time $t$. Transition
probabilities $p_{0\rightarrow1}$ and $p_{1\rightarrow0}$ are unknown
parameters to be estimated from accident severity data. The stationary
unconditional probabilities of states $s_t=0$ and $s_t=1$ are
$\bar p_0=p_{1\rightarrow0}/(p_{0\rightarrow1}+p_{1\rightarrow0})$ and
$\bar p_1=p_{0\rightarrow1}/(p_{0\rightarrow1}+p_{1\rightarrow0})$
respectively.\footnote{These can be found from stationarity conditions
$\bar p_0=(1-p_{0\rightarrow1})\bar p_0+p_{1\rightarrow0}\bar p_1$,
$\bar p_1=p_{0\rightarrow1}\bar p_0+(1-p_{1\rightarrow0})\bar p_1$ and
$\bar p_0+\bar p_1=1$.}
Without loss of generality, we assume that (on average) state $s_t=0$
occurs more or equally frequently than state $s_t=1$. Therefore,
$\bar p_{0}\geq\bar p_{1}$, and we obtain
restriction\footnote{Without any loss of generality,
restriction~(\ref{EQ_P_RESTRICT}) is introduced for the purpose of
avoiding the problem of state label switching \mbox{$0\leftrightarrow1$}.
This problem would otherwise arise because of the symmetry of
Eqs.~(\ref{EQ_P})--(\ref{EQ_L}) under the label switching.}
\begin{eqnarray}
p_{0\rightarrow1}\leq p_{1\rightarrow0}.
\label{EQ_P_RESTRICT}
\end{eqnarray}
We refer to states $s_t=0$ and $s_t=1$ as ``more frequent'' and
``less frequent'' states respectively.

Next, a two-state Markov switching multinomial logit (MSML) model
assumes multinomial logit (ML) data-generating processes for accident
severity in each of the two states. With this, the probability of
the $i^{\rm th}$ severity outcome observed in the $n^{\rm th}$
accident during time period $t$ is
\begin{eqnarray}
P_{t,n}^{(i)} &=& \left\{
  \begin{array}{ll}
  \displaystyle
  \frac{\exp(\betabf_{(0),i}'{\bf X}_{t,n})}
       {\sum_{j=1}^I\exp(\betabf_{(0),j}'{\bf X}_{t,n})}
  & \quad\mbox{if $s_t=0$},\\
  \displaystyle
  \frac{\exp(\betabf_{(1),i}'{\bf X}_{t,n})\vphantom{\int\limits^o}}
       {\sum_{j=1}^I\exp(\betabf_{(1),j}'{\bf X}_{t,n})}
  & \quad\mbox{if $s_t=1$},
\end{array}
\right.
\label{EQ_ML}\\
i &=& 1,2,\ldots,I, \quad n=1,2,\ldots,N_t,
                    \quad t = 1,2,\ldots,T, \vphantom{\int}
\nonumber
\end{eqnarray}
Here prime means transpose (so $\betabf_{(0),i}'$ is the transpose
of $\betabf_{(0),i}$). Parameter vectors $\betabf_{(0),i}$ and
$\betabf_{(1),i}$ are unknown estimable parameters of the
two standard multinomial logit probability mass functions
\citep[][]{WKM_03} in the two states, $s_t=0$ and $s_t=1$
respectively. We set the first component of
${\bf X}_{t,n}$ to unity, and, therefore, the first components of
vectors $\betabf_{(0),i}$ and $\betabf_{(1),i}$ are the intercepts
in the two states. In addition, without loss of generality, we set
all $\beta$-parameters for the last severity outcome to
zero,\footnote{This can be done because ${\bf X}_{t,n}$ are assumed
to be independent of the outcome $i$.}
$\betabf_{(0),I}=\betabf_{(1),I}={\bf 0}$.

If accident events are assumed to be independent, the
likelihood function is
\begin{eqnarray}
f({\bf Y}|{\bf\Theta},{\cal M})=\prod\limits_{t=1}^T
\prod\limits_{n=1}^{N_t} \prod\limits_{i=1}^I
\left[P_{t,n}^{(i)}\right]^{\delta^{(i)}_{t,n}}.
\label{EQ_L}
\end{eqnarray}
Here, because the state variables $s_{t,n}$ are unobservable, the
vector of all estimable parameters ${\bf\Theta}$ must include all
states, in addition to model parameters ($\beta$-s) and transition
probabilities. Thus, ${\bf\Theta}=[\betabf_{(0)}',\betabf_{(1)}',
p_{0\rightarrow1},p_{1\rightarrow0},{\bf S}']'$,
where vector ${\bf S}=[s_1,s_2,...,s_T]'$ has length $T$ and contains
all state values.
Eqs.~(\ref{EQ_P})-(\ref{EQ_L}) define the two-state Markov switching
multinomial logit (MSML) model considered here.

\section{Model estimation methods}
\label{MOD_ESTIM}

Statistical estimation of Markov switching models is complicated by
unobservability of the state variables
$s_t$.\footnote{\label{FN_S}Below we will have 208 time periods
($T=208$). In this case, there are $2^{208}$ possible
combinations for value of vector ${\bf S}=[s_1,s_2,...,s_T]'$.}
As a result, the traditional maximum likelihood estimation (MLE)
procedure is of very limited use for Markov switching models.
Instead, a Bayesian inference approach is used.
Given a model ${\cal M}$ with likelihood function
$f({\bf Y}|{\bf\Theta},{\cal M})$, the Bayes formula is
\begin{eqnarray}
f({\bf\Theta}|{\bf Y},{\cal M})=
\frac{f({\bf Y},{\bf\Theta}|{\cal M})}{f({\bf Y}|{\cal M})}=
\frac{f({\bf Y}|{\bf\Theta},{\cal M})\pi({\bf\Theta}|{\cal M})}
{\int f({\bf Y},{\bf\Theta}|{\cal M})\,d{\bf\Theta}}.
\label{EQ_POSTERIOR}
\end{eqnarray}
Here $f({\bf\Theta}|{\bf Y},{\cal M})$ is the posterior probability
distribution of model parameters ${\bf\Theta}$ conditional on the
observed data ${\bf Y}$ and model ${\cal M}$.
Function $f({\bf Y},{\bf\Theta}|{\cal M})$ is the
joint probability distribution of ${\bf Y}$ and ${\bf\Theta}$ given
model ${\cal M}$. Function $f({\bf Y}|{\cal M})$ is the marginal
likelihood function -- the probability distribution of data
${\bf Y}$ given model ${\cal M}$. Function $\pi({\bf\Theta}|{\cal M})$
is the prior probability distribution of parameters that reflects
prior knowledge about ${\bf\Theta}$. The intuition behind
Eq.~(\ref{EQ_POSTERIOR}) is straightforward: given model ${\cal M}$, the
posterior distribution accounts for both the observations ${\bf Y}$
and our prior knowledge of ${\bf\Theta}$.

In our study (and in most practical studies), the direct application
of Eq.~(\ref{EQ_POSTERIOR}) is not feasible because the parameter vector
${\bf\Theta}$ contains too many components, making integration over
${\bf\Theta}$ in Eq.~(\ref{EQ_POSTERIOR}) extremely difficult. However,
the posterior distribution $f({\bf\Theta}|{\bf Y},{\cal M})$ in
Eq.~(\ref{EQ_POSTERIOR}) is known up to its normalization constant,
$f({\bf\Theta}|{\bf Y},{\cal M})\propto
f({\bf Y}|{\bf\Theta},{\cal M})\pi({\bf\Theta}|{\cal M})$. As
a result, we use Markov Chain Monte Carlo (MCMC) simulations, which
provide a convenient and practical computational methodology for
sampling from a probability distribution known up to a constant (the
posterior distribution in our case). Given a large enough posterior
sample of parameter vector ${\bf\Theta}$, any posterior expectation
and variance can be found and Bayesian inference can be readily applied.
A reader interested in details is referred to our Paper~I or to
\citet{M_08}, where we describe our choice of the prior distribution
$\pi({\bf\Theta}|{\cal M})$ and the MCMC simulation
algorithm.\footnote{Our priors for $\beta$-s, $p_{0\rightarrow1}$ and
$p_{1\rightarrow0}$ are flat or nearly flat, while the prior for
the states ${\bf S}$ reflects the Markov process property, specified
by Eq.~(\ref{EQ_P}).}
Although, in this study we estimate
a two-state Markov switching multinomial logit model for accident
severity outcomes and in Paper~I we estimated a two-state Markov
switching negative binomial model for accident frequencies, this
difference is not essential for the Bayesian-MCMC model estimation
methods. In fact, the main difference is in the likelihood function
(multinomial logit as opposed to negative binomial). So we
used the same our own numerical MCMC code, written in the MATLAB
programming language, for model estimation in both studies.
We tested our code on artificial data sets of accident severity
outcomes. The test procedure included a generation of artificial
data with a known model. Then these data were used to estimate
the underlying model by means of our simulation code. With this
procedure we found that the MSML models, used to generate the
artificial data, were reproduced successfully with our estimation
code.

For comparison of different models we use a formal Bayesian
approach. Let there be two models ${\cal M}_1$ and ${\cal M}_2$
with parameter vectors ${\bf\Theta_1}$ and ${\bf\Theta_2}$
respectively. Assuming that we have equal preferences of these
models, their prior probabilities are
$\pi({\cal M}_1)=\pi({\cal M}_2)=1/2$. In this case, the ratio
of the models' posterior probabilities, $P({\cal M}_1|{\bf Y})$
and $P({\cal M}_2|{\bf Y})$, is equal to the Bayes factor. The
later is defined as the ratio of the models' marginal likelihoods
\citep[see][]{KR_95}. Thus, we have
\begin{eqnarray}
\frac{P({\cal M}_2|{\bf Y})}{P({\cal M}_1|{\bf Y})}=
\frac{f({\cal M}_2,{\bf Y})/f(\bf Y)}{f({\cal M}_1,{\bf Y})/f(\bf Y)}=
\frac{f({\bf Y}|{\cal M}_2)\pi({\cal M}_2)}{f({\bf Y}|{\cal M}_1)\pi({\cal M}_1)}=
\frac{f({\bf Y}|{\cal M}_2)}{f({\bf Y}|{\cal M}_1)},
\label{EQ_BAYES_FACTOR}
\end{eqnarray}
where $f({\cal M}_1,{\bf Y})$ and $f({\cal M}_2,{\bf Y})$ are
the joint distributions of the models and the data, $f({\bf Y})$
is the unconditional distribution of the data. As in Paper~I, to
calculate the marginal likelihoods $f({\bf Y}|{\cal M}_1)$ and
$f({\bf Y}|{\cal M}_2)$, we use the harmonic mean formula
$f({\bf Y}|{\cal M})^{-1}=
E\left[\left.f({\bf Y}|{\bf\Theta},{\cal M})^{-1}\right|{\bf Y}\right]$,
where $E(\ldots|{\bf Y})$ means posterior expectation calculated
by using the posterior distribution. If the
ratio in Eq.~(\ref{EQ_BAYES_FACTOR}) is larger than one, then model
${\cal M}_2$ is favored, if the ratio is less than one,
then model ${\cal M}_1$ is favored. An advantage of the use of
Bayes factors is that it has an inherent penalty for
including too many parameters in the model and guards against
overfitting.

To evaluate the performance of model $\{{\cal M},{\bf\Theta}\}$
in fitting the observed data ${\bf Y}$, we carry out the Pearson's
$\chi^2$ goodness-of-fit test \citep{MS_96,C_98,W_02,PTVF_07}. We
perform this test by Monte Carlo simulations to find the distribution
of the Pearson's $\chi^2$ quantity, which measures the discrepancy
between the observations and the model predictions \citep{C_98}.
This distribution is then used to find the goodness-of-fit p-value,
which is the probability that $\chi^2$ exceeds the observed value
of $\chi^2$ under the hypothesis that the model is true (the
observed value of $\chi^2$ is calculated by using the observed
data ${\bf Y}$). For additional details, please see \citet{M_08}.

\section{Empirical results}
\label{RESULTS}

The severity outcome of an accident is determined by the injury level
sustained by the most injured individual (if any) involved into the
accident. In this study we consider three accident severity outcomes:
``fatality'', ``injury'' and ``PDO (property damage only)'', which we
number as $i=1,2,3$ respectively ($I=3$). We use data
from 811720 accidents that were observed in Indiana in 2003-2006.
As in Paper~I, we use weekly time periods, $t=1,2,3,\ldots,T=208$
in total.\footnote{A week is from Sunday to Saturday, there are 208
full weeks in the 2003-2006 time interval.}
Thus, the state $s_t$ can change every week. To increase the
predictive power of our models, we consider accidents separately for
each combination of accident type (1-vehicle and 2-vehicle) and
roadway class (interstate highways, US routes, state routes, county
roads, streets). We do not consider accidents with more than two
vehicles involved.\footnote{Among 811720 accidents 241011 (29.7\%)
are 1-vehicle, 525035 (64.7\%) are 2-vehicle, and only 45674 (5.6\%)
are accidents with more than two vehicles involved.}
Thus, in total, there are ten roadway-class-accident-type
combinations that we consider. For each roadway-class-accident-type
combination the following three types of accident frequency models
are estimated:
\begin{itemize}
\item
First, we estimate a standard multinomial logit (ML) model without
Markov switching by maximum likelihood estimation
(MLE).\footnote{\label{FN_AIC} To obtain parsimonious standard models,
estimated by MLE, we choose the explanatory variables and their
dummies by using the Akaike Information Criterion (AIC) and the
$5\%$ statistical significance level for the two-tailed t-test.
Minimization of $AIC=2K-2LL$, were $K$ is the number of free
continuous model parameters and $LL$ is the log-likelihood, ensures
an optimal choice of explanatory variables in a model and avoids
overfitting \citep[][]{T_02,WKM_03}. For details on variable
selection, see \citet{M_06}.}
We refer to this model as ``ML-by-MLE''.
\item
Second, we estimate the same standard multinomial logit model by the
Bayesian inference approach and the MCMC simulations. We refer to
this model as ``ML-by-MCMC''. As one expects, the estimated ML-by-MCMC
model turned out to be very similar to the corresponding ML-by-MLE
model (estimated for the same roadway-class-accident-type combination).
\item
Third, we estimate a two-state Markov switching multinomial logit
(MSML) model by the Bayesian-MCMC methods. In order to make comparison
of explanatory variable effects in different models straightforward,
in the MSML model we use only those explanatory variables that enter
the corresponding standard ML
model.\footnote{A formal Bayesian approach to model variable selection
is based on evaluation of model's marginal likelihood and the Bayes
factor~(\ref{EQ_BAYES_FACTOR}). Unfortunately, because MCMC simulations
are computationally expensive, evaluation of marginal likelihoods
for a large number of trial models is not feasible in our study.}
To obtain the final MSML model reported here, we also consecutively
construct and use $60\%$, $85\%$ and $95\%$ Bayesian credible intervals
for evaluation of the statistical significance of each $\beta$-parameter.
As a result, in the final model some components of $\betabf_{(0)}$ and
$\betabf_{(1)}$ are restricted to zero or restricted to be the same in
the two states.\footnote{A $\beta$-parameter is restricted to zero if it
is statistically insignificant. A $\beta$-parameter is restricted to be
the same in the two states if the difference of its values in the
two states is statistically insignificant. A $(1-a)$ credible interval
is chosen in such way that the posterior probabilities of being below
and above it are both equal to $a/2$ (we use significance levels
$a=40\%,15\%,5\%$).}
We refer to this final model as ``MSML''.
\end{itemize}
Note that the two states, and thus the MSML models, do not have to
exist for every roadway-class-accident-type combination. For example,
they will not exist if all estimated model parameters turn out to
be statistically the same in the two states,
$\betabf_{(0)}=\betabf_{(1)}$, (which suggests the two states
are identical and the MSML models reduce to the corresponding
standard ML models). Also, the two states will not exist if all
estimated state variables $s_t$ turn out to be close to zero,
resulting in $p_{0\rightarrow1}\ll p_{1\rightarrow0}$ [compare
to Eq.~(\ref{EQ_P_RESTRICT})], then the less frequent state
$s_t=1$ is not realized and the process stays in state $s_t=0$.

Turning to the estimation results, the findings show that two
states of roadway safety and the appropriate MSML models exist
for severity outcomes of 1-vehicle accidents occurring on all
roadway classes (interstate highways, US routes, state routes,
county roads, streets), and for severity outcomes of 2-vehicle
accidents occurring on streets. We did not find two states in the
cases of 2-vehicle accidents on interstate highways, US routes,
state routes and county roads (in these cases all estimated state
variables $s_t$ were found to be close to zero). The model
estimation results for severity outcomes of 1-vehicle accidents
occurring on interstate highways, US routes and state routes are
given in Tables~\ref{T_1}--\ref{T_3}. All continuous model
parameters ($\beta$-s, $p_{0\rightarrow1}$ and $p_{1\rightarrow0}$)
are given together with their $95\%$ confidence intervals (if MLE)
or $95\%$ credible intervals (if Bayesian-MCMC), refer to the
superscript and subscript numbers adjacent to parameter estimates in
Tables~\ref{T_1}--\ref{T_3}.\footnote{Note that MLE assumes
asymptotic normality of the estimates, resulting in confidence
intervals being symmetric around the means (a $95\%$ confidence
interval is $\pm1.96$ standard deviations around the mean). In
contrast, Bayesian estimation does not require this assumption, and
posterior distributions of parameters and Bayesian credible intervals
are usually non-symmetric.}
Table~\ref{T_4} gives summary statistics of all roadway accident
characteristic variables ${\bf X}_{t,n}$ (except the intercept).

\begin{landscape}
\begin{table}[p]
\caption{Estimation results for multinomial logit models of severity
outcomes of one-vehicle accidents on Indiana interstate highways}
{(the superscript and subscript numbers to the right of individual
parameter estimates are $95\%$ confidence/credible intervals)}
\label{T_1}
\begin{scriptsize}
\input{ms_3_Table1a.tex}

\end{scriptsize}
\end{table}
\addtocounter{table}{-1}
\begin{table}[p]
\caption{(Continued)}
\begin{scriptsize}

\input{ms_3_Table1b.tex}

\end{scriptsize}
\end{table}
\end{landscape}

\begin{landscape}
\begin{table}[p]
\caption{Estimation results for multinomial logit models of severity
outcomes of one-vehicle accidents on Indiana US routes}
{(the superscript and subscript numbers to the right of individual
parameter estimates are $95\%$ confidence/credible intervals)}
\label{T_2}
\begin{scriptsize}
\input{ms_3_Table2a.tex}

\end{scriptsize}
\end{table}
\addtocounter{table}{-1}
\begin{table}[p]
\caption{(Continued)}
\begin{scriptsize}

\input{ms_3_Table2b.tex}

\end{scriptsize}
\end{table}
\end{landscape}

\begin{landscape}
\begin{table}[p]
\caption{Estimation results for multinomial logit models of severity
outcomes of one-vehicle accidents on Indiana state routes}
{(the superscript and subscript numbers to the right of individual
parameter estimates are $95\%$ confidence/credible intervals)}
\label{T_3}
\begin{scriptsize}
\input{ms_3_Table3a.tex}

\end{scriptsize}
\end{table}
\addtocounter{table}{-1}
\begin{table}[p]
\caption{(Continued)}
\begin{scriptsize}

\input{ms_3_Table3b.tex}

\end{scriptsize}
\end{table}
\end{landscape}

\begin{table}[p]
\caption{Summary statistics of roadway accident characteristic variables}
\label{T_4}
\begin{scriptsize}
\input{ms_3_Table4a.tex}

\end{scriptsize}
\end{table}
\addtocounter{table}{-1}
\begin{table}[p]
\caption{(Continued)}
\begin{scriptsize}
\input{ms_3_Table4b.tex}

\end{scriptsize}
\end{table}

\begin{figure}[t]
\vspace{14.1truecm}
\includegraphics{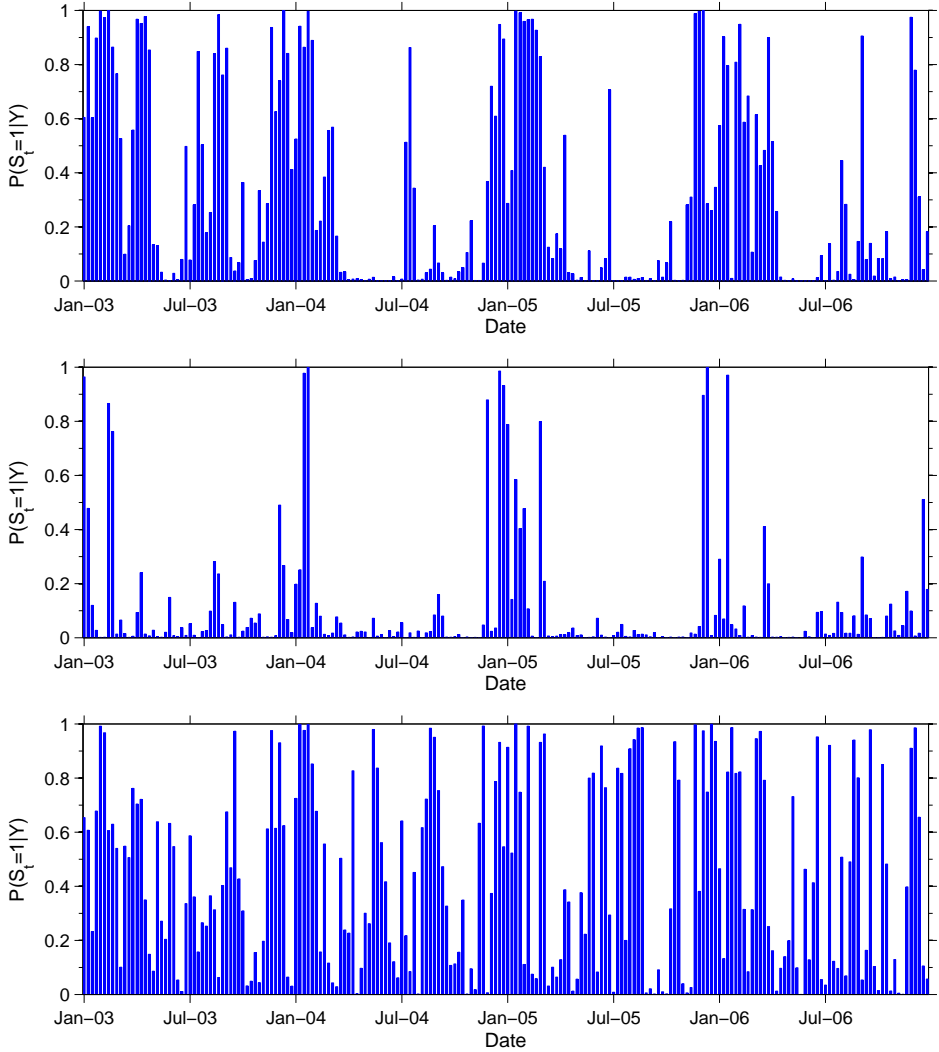}
\caption{Weekly posterior probabilities $P(s_t=1|{\bf Y})$
for the MSML models estimated for severity of 1-vehicle accidents
on interstate highways (top plot), US routes (middle plot) and
state routes (bottom plot).
\label{FIGURE}
}
\end{figure}

The top, middle and bottom plots in Figure~\ref{FIGURE} show weekly
posterior probabilities $P(s_t=1|{\bf Y})$ of the less frequent state
$s_t=1$ for the MSML models estimated for severity of 1-vehicle
accidents occurring on interstate highways, US routes and state
routes respectively.\footnote{Note that these posterior
probabilities are equal to the posterior expectations of $s_t$,
$P(s_t=1|{\bf Y})=1\times P(s_t=1|{\bf Y})+0\times P(s_t=0|{\bf Y})=E(s_t|{\bf Y})$.}
Because of space limitations, in this paper we do not report
estimation results for severity of 1-vehicle accidents on county
roads and streets, and for severity of 2-vehicle accidents. However,
below we discuss our findings for all roadway-class-accident-type
combinations. For unreported model estimation results see
\citet[][]{M_08}.

We find that in all cases when the two states and Markov switching
multinomial logit (MSML) models exist, these models are strongly
favored by the empirical data over the corresponding standard
multinomial logit (ML) models. Indeed, from lines ``marginal $LL$''
in Tables~\ref{T_1}--\ref{T_3} we see that the MSML models provide
considerable, ranging from $40.5$ to $61.4$, improvements of the
logarithm of the marginal likelihood of the data as compared to the
corresponding ML models.\footnote{We use the harmonic mean formula
to calculate the values and the $95\%$ confidence intervals of the
log-marginal-likelihoods given in lines ``marginal $LL$'' of
Tables~\ref{T_1}--\ref{T_3}. The confidence intervals are calculated
by bootstrap simulations. For details, see Paper~I or \citet{M_08}.}
Thus, from Eq.~(\ref{EQ_BAYES_FACTOR}) we find that, given the accident
severity data, the posterior probabilities of the MSML models are larger
than the probabilities of the corresponding ML models by factors
ranging from $e^{40.5}$ to $e^{61.4}$. In the cases of 1-vehicle
accidents on county roads, streets and the case of 2-vehicle accidents
on streets, MSML models (not reported here) are also strongly favored
by the empirical data over the corresponding ML models
\citep[][]{M_08}.

Let us now consider the maximum likelihood estimation (MLE)
of the standard ML models and an imaginary MLE estimation of the
MSML models. We find that, in this imaginary case, a classical
statistics approach for model comparison, based on the MLE, would
also favors the MSML models over the standard ML models. For example,
refer to line ``max$(LL)$'' in Table~\ref{T_1} given for the case
of 1-vehicle accidents on interstate highways.
The MLE gave the maximum log-likelihood value $-8465.79$ for the
standard ML model. The maximum log-likelihood value observed during
our MCMC simulations for the MSML model is equal to $-8358.97$.
An imaginary MLE, at its convergence, would give a MSML log-likelihood
value that would be even larger than this observed value. Therefore,
if estimated by the MLE, the MSML model would provide large, at least
$106.82$ improvement in the maximum log-likelihood value over the
corresponding ML model. This improvement would come with only modest
increase in the number of free continuous model parameters ($\beta$-s)
that enter the likelihood function (refer to Table~\ref{T_1} under
``\# free par.''). Similar arguments hold for comparison of MSML
and ML models estimated for other roadway-class-accident-type
combinations (see Tables~\ref{T_2} and~\ref{T_3}).

To evaluate the goodness-of-fit for a model, we use the posterior
(or MLE) estimates of all continuous model parameters ($\beta$-s,
$\alpha$, $p_{0\rightarrow1}$, $p_{1\rightarrow0}$)
and generate $10^4$ artificial data sets under the hypothesis
that the model is true.\footnote{Note that the state values $\bf S$
are generated by using $p_{0\rightarrow1}$ and
$p_{1\rightarrow0}$.}
We find the distribution of $\chi^2$ and calculate the
goodness-of-fit p-value for the observed value of $\chi^2$.
For details, see \citet{M_08}. The resulting p-values for
our models are given in Tables~\ref{T_1}--\ref{T_3}. These p-values
are around \mbox{$00$--$100\%$}. Therefore, all models fit the data
well.

Now, refer to Table~\ref{T_5}. The first six rows of this table
list time-correlation coefficients between posterior probabilities
$P(s_t=1|{\bf Y})$ for the six MSML models that exist and are
estimated for six roadway-class-accident-type combinations (1-vehicle
accidents on interstate highways, US routes, state routes, county
roads, streets, and 2-vehicle accidents on streets).\footnote{Here
and below we calculate weighted correlation coefficients. For variable
$P(s_t=1|{\bf Y})\equiv E(s_t|{\bf Y})$ we use weights $w_t$ inversely
proportional to the posterior standard deviations of $s_t$. That is
$w_t\propto \min\left\{1/{\rm std}(s_t|{\bf Y}),
{\rm median}[1/{\rm std}(s_t|{\bf Y})]\right\}$.}
We see that the states for 1-vehicle accidents
on all high-speed roads (interstate highways, US routes, state routes
and county roads) are correlated with each other. The values of the
corresponding correlation coefficients are positive and range from
$0.263$ to $0.688$ (see Table~\ref{T_5}). This result suggests an
existence of common (unobservable) factors that can cause switching
between states of roadway safety for 1-vehicle accidents on all
high-speed roads.

\begin{table}[t]
\caption{Correlations of the posterior probabilities $P(s_t=1|{\bf Y})$
with each other and with weather-condition variables (for the MSML model)}
\label{T_5}
\begin{scriptsize}

\input{ms_3_Table5.tex}

\end{scriptsize}
\end{table}

The remaining rows of Table~\ref{T_5} show correlation coefficients
between posterior probabilities $P(s_t=1|{\bf Y})$ and weather-condition
variables. These correlations were found by using daily and hourly
historical weather data in Indiana, available at the Indiana State
Climate Office at Purdue University (www.agry.purdue.edu/climate).
For these correlations, the precipitation and snowfall amounts are
daily amounts in inches averaged over the week and across Indiana
weather observation stations.\footnote{Snowfall
and precipitation amounts are weakly related with each other because
snow density $(g/cm^3)$ can vary by more than a factor of ten.}
The temperature variable is the mean daily air temperature $(^oF)$
averaged over the week and across the weather stations.
The wind gust variable is the maximal instantaneous wind speed (mph)
measured during the 10-minute period just prior to the observational
time. Wind gusts are measured every hour and averaged over the week
and across the weather stations.
The effect of fog/frost is captured by a dummy variable that is equal
to one if and only if the difference between air and dewpoint
temperatures does not exceed $5^oF$ (in this case frost can form if
the dewpoint is below the freezing point $32^oF$, and fog can form
otherwise). The fog/frost dummies are calculated for every hour and
are averaged over the week and across the weather stations.
Finally, visibility distance variable is the harmonic mean of hourly
visibility distances, which are measured in miles every hour and are
averaged over the week and across the weather stations.\footnote{The
harmonic mean $\bar d$ of distances $d_n$ is calculated as $\bar d^{-1}=(1/N)\sum^N_{n=1}d^{-1}_n$, assuming $d_n=0.25$ miles if
$d_n\leq0.25$ miles.}

From the results given in Table~\ref{T_5} we find that for 1-vehicle
accidents  on all high-speed roads (interstate highways, US routes,
state routes and county roads), the less frequent state $s_t=1$ is
positively correlated with extreme temperatures (low during winter
and high during summer), rain precipitations and snowfalls, strong
wind gusts, fogs and frosts, low visibility distances. It is
reasonable to expect that roadway safety is different during
bad weather as compared to better weather, resulting in the
two-state nature of roadway safety.

The results of Table~\ref{T_5} suggest that Markov switching for
road safety on streets is very different from switching on all other
roadway classes. In particular, the states of roadway safety
on streets exhibit low correlation with states on
other roads. In addition, only streets exhibit Markov switching in
 the case of 2-vehicle accidents. Finally, states of roadway safety
on streets show little correlation with weather conditions. A
possible explanation of these differences is that streets are
mostly located in urban areas and they have traffic moving at
speeds lower that those on other roads.

Next, we consider the estimation results for the stationary
unconditional probabilities $\bar p_0$ and $\bar p_1$ of states
$s_t=0$ and $s_t=1$ for MSML models (see Section~\ref{MOD_SPECIF}).
In the cases of 1-vehicle accidents on interstate highways,
US routes and state routes these transition probabilities are
listed in lines ``${\bar p}_0$ and ${\bar p}_1$'' of
Tables~\ref{T_1}--\ref{T_3}. In the cases of 1-vehicle accidents
on county roads and 1- and 2-vehicle accidents on streets
refer to \citet[][]{M_08}. We find that the ratio
$\bar p_1/\bar p_0$ is approximately equal to
$0.46$, $0.13$, $0.74$, $0.25$, $0.65$ and $0.36$
in the cases of 1-vehicle accidents on interstate highways, US
routes, state routes, county roads, streets, and 2-vehicle
accidents on streets respectively.
Thus for some roadway-class-accident-type combinations (for example,
1-vehicle accidents on US routes) the less frequent state $s_t=1$
is quite rare, while for other combinations (for example, 1-vehicle
accidents on state routes) state $s_t=1$ is only slightly less
frequent than state $s_t=0$.

Finally, we set model parameters ($\beta$-s) to their posterior
means, calculate the probabilities of fatality and injury outcomes
by using Eq.~(\ref{EQ_ML}) and average these probabilities over
all values of the explanatory variables ${\bf X}_{t,n}$ observed in
the data sample. We compare these probabilities across the two states
of roadway safety, $s_t=0$ and $s_t=1$, for MSML models [refer to
lines ``$\langle P_{t,n}^{(i)}\rangle_X$'' in
Tables~\ref{T_1}--\ref{T_3} and to \citet[][]{M_08}]. We find that
in many cases these averaged probabilities of fatality and injury
outcomes do not differ very significantly across the two states of
roadway safety (the only significant differences are for fatality
probabilities in the cases of 1-vehicle accidents on US routes,
county roads and streets). This means that in many cases states
$s_t=0$ and $s_t=1$ are approximately equally dangerous as far as
accident severity is concerned. We discuss this result in the next
section.

\section{Conclusions}
\label{CONCLUD}

In this study we found that two states of roadway safety and Markov
switching multinomial logit (MSML) models exist for severity of
1-vehicle accidents occurring on high-speed roads (interstate
highways, US routes, state routes, county roads), but not for
2-vehicle accidents on high-speed roads. One of possible explanations
of this result is that 1- and 2-vehicle accidents may differ in their
nature. For example, on one hand, severity of 1-vehicle accidents may
frequently be determined by driver-related factors (speeding, falling
a sleep, driving under the influence, etc). Drivers' behavior might
exhibit a two-state pattern. In particular, drivers might be
overconfident and/or have difficulties in adjustments to bad weather
conditions. On the other hand, severity of a 2-vehicle accident might
crucially depend on the actual physics involved in the collision
between the two cars (for example, head-on and
side impacts are more dangerous than rear-end collisions). As far as
slow-speed streets are concerned, in this case both 1- and 2-vehicle
accidents exhibit two-state nature for their severity. Further
studies are needed to understand these results. In this study, the
important result is that in all cases when two states of roadway
safety exist, the two-state MSML models provide much superior
statistical fit for accident severity outcomes as compared to
the standard ML models.

We found that in many cases states $s_t=0$ and $s_t=1$ are
approximately equally dangerous as far as accident severity is
concerned. This result holds despite the fact that state $s_t=1$
is correlated with adverse weather conditions. A likely and simple
explanation of this finding is that during bad weather both number
of serious accidents (fatalities and injuries) and number of minor
accidents (PDOs) increase, so that their relative fraction stays
approximately steady. In addition, most drivers are rational and
they are likely take some precautions while driving during
bad weather. From the results presented in Paper~I we know
that the total number of accidents significantly increases during
adverse weather conditions. Thus, driver's precautions are probably
not sufficient to avoid increases in accident rates during bad
weather.



\end{document}

%% file: ms_3_Table1a.tex
\tabcolsep=0.3em
\renewcommand{\arraystretch}{1.45}

\begin{tabular}{|l|c|c|c|c|c|c|c|c|}
\hline
& \multicolumn{2}{|c|}{} & \multicolumn{2}{|c|}{} &
\multicolumn{4}{|c|}{\bf MSML$^{\,\rm c}$}
\\
\cline{6-9}
${}^{{}^{\bf\displaystyle Variable}}$ &
\multicolumn{2}{|c|}{${}^{{}^{\bf\displaystyle ML\mbox{\bf-}by\mbox{\bf-}MLE^{\,\rm a}}}$} & \multicolumn{2}{|c|}{${}^{{}^{\bf\displaystyle ML\mbox{\bf-}by\mbox{\bf-}MCMC^{\,\rm b}}}$} &
\multicolumn{2}{|c|}{{\bf state }{\boldmath$s=0$}} &
\multicolumn{2}{|c|}{{\bf state }{\boldmath$s=1$}}
\\
\cline{2-9}
& {\bf fatality} & {\bf injury} & {\bf fatality} & {\bf injury}
& {\bf fatality} & {\bf injury} & {\bf fatality} & {\bf injury}
\\
\hline
Intercept (constant term) & $-11.9^{-10.1}_{-13.7}$ &
$-3.69^{-3.53}_{-3.84}$ &
$-12.4^{-10.6}_{-14.5}$ &
$-3.72^{-3.56}_{-3.88}$ &
$-12.2^{-10.5}_{-14.4}$ &
$-3.98^{-3.79}_{-4.17}$ &
$-12.2^{-10.5}_{-14.4}$ &
$-3.22^{-2.98}_{-3.45}$\\
\hline
Summer season (dummy) & $.235^{.329}_{.142}$ &
$.235^{.329}_{.142}$ &
$.237^{.329}_{.143}$ &
$.237^{.329}_{.143}$ &
$.176^{.293}_{.0551}$ &
$.176^{.293}_{.0551}$ &
$.176^{.293}_{.0551}$ &
$.615^{.959}_{.282}$\\
\hline
Thursday (dummy) &
$-.798^{-.115}_{-1.48}$ &
-- &
$-.853^{-.206}_{-1.59}$ &
-- &
$-.872^{-.225}_{-1.61}$ &
-- &
$-.872^{-.225}_{-1.61}$ &
--\\
\hline
Construction at the accident location (dummy) &
$-.418^{-.213}_{-.623}$ &
$-.418^{-.213}_{-.623}$ &
$-.425^{-.224}_{-.632}$ &
$-.425^{-.224}_{-.632}$ &
$-.566^{-.319}_{-.822}$ &
$-.566^{-.319}_{-.822}$ &
$-.566^{-.319}_{-.822}$ &
--\\
\hline
Daylight or street lights are lit up if dark (dummy) &
$-.392^{-.0368}_{-.748}$ &
$.137^{.224}_{.0501}$ &
$-.387^{-.0301}_{-.740}$ &
$.143^{.230}_{.0568}$ &
$-.378^{-.0236}_{-.729}$ &
$.139^{.226}_{.0522}$ &
$-.378^{-.0236}_{-.729}$ &
$.139^{.226}_{.0522}$\\
\hline
Precipitation: rain/freezing rain/snow/sleet/hail (dummy) &
$-1.38^{-.830}_{-1.92}$ &
$-.361^{-.264}_{-.457}$ &
$-1.41^{-.884}_{-1.99}$ &
$-.363^{-.267}_{-.460}$ &
$-1.54^{-1.03}_{-2.10}$ &
$-.563^{-.404}_{-.729}$ &
$-1.54^{-1.03}_{-2.10}$ &
--\\
\hline
Roadway surface is covered by snow/slush (dummy) &
$-1.28^{-.0917}_{-2.46}$ &
$-.432^{-.280}_{-.583}$ &
$-1.43^{-.328}_{-2.84}$ &
$-.438^{-.288}_{-.590}$ &
$-.0515^{-.361}_{-.671}$ &
$-.0515^{-.361}_{-.671}$ &
$-.0515^{-.361}_{-.671}$ &
$-.0515^{-.361}_{-.671}$\\
\hline
Roadway median is drivable (dummy) &
$.571^{.929}_{.213}$ &
-- &
$.577^{.939}_{.223}$ &
-- &
$.566^{.930}_{.211}$ &
-- &
$.566^{.930}_{.211}$ &
--\\
\hline
Roadway is at curve (dummy) &
$.114^{.212}_{.0165}$ &
$.114^{.212}_{.0165}$ &
$.116^{.213}_{.0186}$ &
$.116^{.213}_{.0186}$ &
-- &
-- &
-- &
--\\
\hline
Primary cause of the accident is driver-related (dummy) &
$4.24^{5.30}_{3.18}$ &
$1.53^{1.64}_{1.43}$ &
$4.39^{5.64}_{3.39}$ &
$1.54^{1.64}_{1.43}$ &
$4.48^{5.73}_{3.48}$ &
$2.00^{2.18}_{1.84}$ &
$4.48^{5.73}_{3.48}$ &
$.715^{.946}_{.468}$\\
\hline
Help arrived in 20 minutes or less after the crash (dummy) &
$.790^{.887}_{.693}$ &
$.790^{.887}_{.693}$ &
$.790^{.891}_{.691}$ &
$.790^{.891}_{.691}$ &
$.785^{.886}_{.684}$ &
$.785^{.886}_{.684}$ &
$.785^{.886}_{.684}$ &
$.785^{.886}_{.684}$\\
\hline
The vehicle at fault is a motorcycle (dummy) &
$3.88^{4.59}_{3.17}$ &
$2.74^{3.12}_{2.36}$ &
$3.87^{4.57}_{3.13}$ &
$2.75^{3.15}_{2.37}$ &
$4.61^{5.49}_{3.74}$ &
$3.23^{3.83}_{2.70}$ &
-- &
$1.39^{2.49}_{.326}$\\
\hline
Age of the vehicle at fault (in years) &
$.0285^{.0370}_{.0201}$ &
$.0285^{.0370}_{.0201}$ &
$.0286^{.0370}_{.0201}$ &
$.0286^{.0370}_{.0201}$ &
-- &
$.0286^{.0371}_{.0200}$ &
-- &
$.0286^{.0371}_{.0200}$\\
\hline
Number of occupants in the vehicle at fault &
$.366^{.463}_{.269}$ &
$.123^{.159}_{.0859}$ &
$.367^{.465}_{.264}$ &
$.123^{.159}_{.0861}$ &
$.366^{.464}_{.263}$ &
$.124^{.161}_{.0874}$ &
$.366^{.464}_{.263}$ &
$.124^{.161}_{.0874}$\\
\hline
Roadway traveled by the vehicle at fault is multi-lane and
& & & & & & & &\\
divided two-way (dummy) &
$2.60^{4.00}_{1.20}$ &
-- &
$2.86^{4.63}_{1.56}$ &
-- &
$2.86^{4.66}_{1.56}$ &
-- &
$2.86^{4.66}_{1.56}$ &
--\\
\hline
At least one of the vehicles involved was on fire (dummy) &
$1.24^{2.12}_{}$ &
$-.345^{-.0257}_{-.665}$ &
$1.18^{2.02}_{.206}$ &
$-.345^{-.0335}_{-.669}$ &
$1.66^{2.56}_{.621}$ &
$-.332^{-.0198}_{-.659}$ &
-- &
$-.332^{-.0198}_{-.659}$\\
\hline
Gender of the driver at fault (dummy) & -- &
$.328^{.410}_{.246}$ &
-- &
$.331^{.413}_{.248}$ &
-- &
$.224^{.338}_{.107}$ &
-- &
$.479^{.637}_{.328}$\\
\hline
\end{tabular}

%% file: ms_3_Table1b.tex
\renewcommand{\arraystretch}{1.45}

\begin{tabular}{|l|c|c|c|c|c|c|c|c|}
\hline
& \multicolumn{2}{|c|}{} & \multicolumn{2}{|c|}{} &
\multicolumn{4}{|c|}{\bf MSML$^{\,\rm c}$}
\\
\cline{6-9}
${}^{{}^{\bf\displaystyle Variable}}$ &
\multicolumn{2}{|c|}{${}^{{}^{\bf\displaystyle ML\mbox{\bf-}by\mbox{\bf-}MLE^{\,\rm a}}}$} & \multicolumn{2}{|c|}{${}^{{}^{\bf\displaystyle ML\mbox{\bf-}by\mbox{\bf-}MCMC^{\,\rm b}}}$} &
\multicolumn{2}{|c|}{{\bf state }{\boldmath$s=0$}} &
\multicolumn{2}{|c|}{{\bf state }{\boldmath$s=1$}}
\\
\cline{2-9}
& {\bf fatality} & {\bf injury} & {\bf fatality} & {\bf injury}
& {\bf fatality} & {\bf injury} & {\bf fatality} & {\bf injury}
\\
\hline
Probability of severity outcome [$P_{t,n}^{(i)}$
given by Eq.~(\ref{EQ_ML})], averaged
& & & & & & & &\\
over all values of explanatory variables ${\bf X}_{t,n}$ & -- &
-- &
$.00724$ &
$.176$ &
$.00733$ &
$.174$ &
$.00672$ &
$.192$\\
\hline
Markov transition probability of jump $0\to1$ ($p_{0\rightarrow1}$) & \multicolumn{2}{|c|}{--} &
\multicolumn{2}{|c|}{--} &
\multicolumn{4}{|c|}{$.151^{.254}_{.0704}$}\\
\hline
Markov transition probability of jump $1\to0$ ($p_{1\rightarrow0}$) & \multicolumn{2}{|c|}{--} &
\multicolumn{2}{|c|}{--} &
\multicolumn{4}{|c|}{$.330^{.532}_{.164}$}\\
\hline
Unconditional probabilities of states 0 and 1 (${\bar p}_0$ and ${\bar p}_1$) &
\multicolumn{2}{|c|}{--} &
\multicolumn{2}{|c|}{--} &
\multicolumn{4}{|c|}{$.683^{.814}_{.540}$\quad and\quad $.317^{.460}_{.186}$}
\\
\hline
\hline
Total number of free model parameters ($\beta$-s) &
\multicolumn{2}{|c|}{$25$} &
\multicolumn{2}{|c|}{$25$} &
\multicolumn{4}{|c|}{$28$}\\
\hline
Posterior average of the log-likelihood (LL) &
\multicolumn{2}{|c|}{--} &
\multicolumn{2}{|c|}{$-8486.78^{-8480.82}_{-8494.61}$} &
\multicolumn{4}{|c|}{$-8396.78^{-8379.21}_{-8416.57}$}\\
\hline
Max$(LL)$:\quad estimated max.~log-likelihood (LL) for MLE; &
\multicolumn{2}{|c|}{} & \multicolumn{2}{|c|}{} &
\multicolumn{4}{|c|}{} \\
maximum observed value of LL for Bayesian-MCMC & \multicolumn{2}{|c|}{$-8465.79\,{\rm(MLE)}\!$} &
\multicolumn{2}{|c|}{$-8476.37\,{\rm(observed)}\!$} &
\multicolumn{4}{|c|}{$-8358.97\,{\rm(observed)}\!$}\\
\hline
Logarithm of marginal likelihood of data ($\ln[f({\bf Y}|{\cal M})]$) &
\multicolumn{2}{|c|}{--} &
\multicolumn{2}{|c|}{$-8498.46^{-8494.22}_{-8499.21}$} &
\multicolumn{4}{|c|}{$-8437.07^{-8424.77}_{-8440.02}$}\\
\hline
Goodness-of-fit p-value & \multicolumn{2}{|c|}{--} &
\multicolumn{2}{|c|}{$0.255$} &
\multicolumn{4}{|c|}{$0.222$}\\
\hline
Maximum of the potential scale reduction factors (PSRF)$^{\,\rm d}$ &
\multicolumn{2}{|c|}{--} &
\multicolumn{2}{|c|}{$1.00302$} &
\multicolumn{4}{|c|}{$1.00060$}\\
\hline
Multivariate potential scale reduction factor (MPSRF)$^{\,\rm d}$ &
\multicolumn{2}{|c|}{--} &
\multicolumn{2}{|c|}{$1.00325$} &
\multicolumn{4}{|c|}{$1.00067$}\\
\hline
Number of available observations & \multicolumn{8}{|c|}
{accidents = fatalities + injuries + PDOs:\quad\quad\quad
$19094=143+3369+15582$}\\
\hline
\multicolumn{9}{l}{$^{\rm a}$ Standard (conventional)
multinomial logit (ML) model estimated by maximum likelihood
estimation (MLE).}\\
\multicolumn{9}{l}{$^{\rm b}$ Standard multinomial logit (ML)
model estimated by Markov Chain Monte Carlo (MCMC) simulations.}\\
\multicolumn{9}{l}{$^{\rm c}$ Two-state Markov switching
multinomial logit (MSML) model estimated by Markov Chain Monte Carlo
(MCMC) simulations.}\\
\multicolumn{9}{l}{$^{\rm d}$ PSRF/MPSRF are calculated separately/jointly
for all continuous model parameters. PSRF and MPSRF are close to 1 for
converged MCMC chains.}\\
\end{tabular}

%% file: ms_3_Table2a.tex
\tabcolsep=0.3em

\begin{tabular}{|l|c|c|c|c|c|c|c|c|}
\hline
& \multicolumn{2}{|c|}{} & \multicolumn{2}{|c|}{} &
\multicolumn{4}{|c|}{\bf MSML$^{\,\rm c}$}
\\
\cline{6-9}
${}^{{}^{\bf\displaystyle Variable}}$ &
\multicolumn{2}{|c|}{${}^{{}^{\bf\displaystyle ML\mbox{\bf-}by\mbox{\bf-}MLE^{\,\rm a}}}$} & \multicolumn{2}{|c|}{${}^{{}^{\bf\displaystyle ML\mbox{\bf-}by\mbox{\bf-}MCMC^{\,\rm b}}}$} &
\multicolumn{2}{|c|}{{\bf state }{\boldmath$s=0$}} &
\multicolumn{2}{|c|}{{\bf state }{\boldmath$s=1$}}
\\
\cline{2-9}
& {\bf fatality} & {\bf injury} & {\bf fatality} & {\bf injury}
& {\bf fatality} & {\bf injury} & {\bf fatality} & {\bf injury}
\\
\hline
Intercept (constant term) & $-6.51^{-5.00}_{-8.03}$ &
$-2.13^{-1.79}_{-2.47}$ &
$-6.62^{-5.16}_{-8.14}$ &
$-2.12^{-1.78}_{-2.47}$ &
$-5.72^{-4.69}_{-6.92}$ &
$-2.05^{-1.71}_{-2.40}$ &
$-5.72^{-4.69}_{-6.92}$ &
$-2.79^{-2.37}_{-3.23}$\\
\hline
Summer season (dummy) & $.514^{.894}_{.134}$ &
$.200^{.305}_{.0947}$ &
$.509^{.883}_{.124}$ &
$.200^{.305}_{.0951}$ &
$.190^{.300}_{.0789}$ &
$.190^{.300}_{.0789}$ &
$.190^{.300}_{.0789}$ &
--\\
\hline
Daylight or street lights are lit up if dark (dummy) &
$-.498^{-.142}_{-.855}$ &
$.194^{.287}_{.101}$ &
$-.492^{-.136}_{-.848}$ &
$.203^{.296}_{.110}$ &
$-.493^{-.136}_{-.857}$ &
$.197^{.290}_{.105}$ &
-- &
$.197^{.290}_{.105}$\\
\hline
Snowing weather (dummy) &
$-1.17^{-.170}_{-2.18}$ &
-- &
$-1.30^{-.357}_{-2.47}$ &
-- &
$-1.10^{-.151}_{-2.27}$ &
$.165^{.317}_{.0115}$ &
$-1.10^{-.151}_{-2.27}$ &
$.165^{.317}_{.0115}$\\
\hline
No roadway junction at the accident location (dummy) &
$.701^{1.25}_{.149}$ &
$.217^{.335}_{.0994}$ &
$.727^{1.31}_{.199}$ &
$.213^{.331}_{.0968}$ &
$.787^{1.36}_{.259}$ &
$.214^{.332}_{.0965}$ &
$.787^{1.36}_{.259}$ &
$.214^{.332}_{.0965}$\\
\hline
Roadway is straight (dummy) &
$-.741^{-.383}_{-1.10}$ &
$-.295^{-.191}_{-.399}$ &
$-.739^{-.377}_{-1.09}$ &
$-.296^{-.192}_{-.399}$ &
$-7.37^{-.372}_{-1.09}$ &
$-.294^{-.189}_{-.398}$ &
$-7.37^{-.372}_{-1.09}$ &
$-.294^{-.189}_{-.398}$\\
\hline
Primary cause of the accident is environment-related (dummy) &
$-3.45^{-2.72}_{-4.18}$ &
$-1.89^{-1.78}_{-1.99}$ &
$-3.51^{-2.81}_{-4.32}$ &
$-1.89^{-1.79}_{-2.00}$ &
$-3.59^{-2.89}_{-4.40}$ &
$-2.09^{-1.96}_{-2.24}$ &
$-3.59^{-2.89}_{-4.40}$ &
$-.701^{-.263}_{-1.16}$\\
\hline
Help arrived in 10 minutes or less after the crash (dummy) &
$.594^{.681}_{.507}$ &
$.594^{.681}_{.507}$ &
$.562^{.650}_{.475}$ &
$.562^{.650}_{.475}$ &
$.560^{.648}_{.472}$ &
$.560^{.648}_{.472}$ &
$.560^{.648}_{.472}$ &
$.560^{.648}_{.472}$\\
\hline
The vehicle at fault is a motorcycle (dummy) &
$2.62^{3.47}_{1.78}$ &
$3.20^{3.55}_{2.86}$ &
$2.57^{3.38}_{1.65}$ &
$3.21^{3.56}_{2.87}$ &
$3.22^{3.58}_{2.88}$ &
$3.22^{3.58}_{2.88}$ &
$3.22^{3.58}_{2.88}$ &
$3.22^{3.58}_{2.88}$\\
\hline
Age of the vehicle at fault (in years) &
$.0363^{.0444}_{.0283}$ &
$.0363^{.0444}_{.0283}$ &
$.0367^{.0448}_{.0287}$ &
$.0367^{.0448}_{.0287}$ &
-- &
$.0366^{.0447}_{.0285}$ &
-- &
$.0366^{.0447}_{.0285}$\\
\hline
Speed limit (used if known and the same for all vehicles involved) &
$.0363^{.0631}_{.00950}$ &
$.0121^{.0178}_{.00640}$ &
$.0373^{.0643}_{.0117}$ &
$.0118^{.0176}_{.00616}$ &
$.0285^{.0495}_{.0104}$ &
$.0102^{.0178}_{.00635}$ &
-- &
$.0120^{.0178}_{.00635}$\\
\hline
Roadway traveled by the vehicle at fault is two-lane and
& & & & & & & &\\
one-way (dummy) &
$-.216^{.0417}_{-.391}$ &
$-.216^{.0417}_{-.391}$ &
$-.223^{.0517}_{-.398}$ &
$-.223^{.0517}_{-.398}$ &
$-.224^{.0504}_{-.401}$ &
$-.224^{.0504}_{-.401}$ &
$-.224^{.0504}_{-.401}$ &
$-.224^{.0504}_{-.401}$\\
\hline
At least one of the vehicles involved was on fire (dummy) &
$1.19^{1.94}_{.439}$ &
-- &
$1.13^{1.85}_{.315}$ &
-- &
$1.27^{1.98}_{.452}$ &
-- &
$1.27^{1.98}_{.452}$ &
--\\
\hline
Age of the driver at fault (in years) &
$.0114^{.0213}_{.00150}$ &
-- &
$.0113^{.0211}_{.00137}$ &
-- &
$.0101^{.0200}_{.0000542}$ &
-- &
-- &
--\\
\hline
Weekday (Monday through Friday) (dummy) & -- &
$-.104^{.0116}_{-.196}$ &
-- &
$-.104^{.0124}_{-.196}$ &
-- &
$-.125^{.0242}_{-.227}$ &
-- &
--\\
\hline
Gender of the driver at fault (dummy) & -- &
$.272^{.362}_{.183}$ &
-- &
$.276^{.365}_{.186}$ &
-- &
$.280^{.369}_{.190}$ &
-- &
$.280^{.369}_{.190}$\\
\hline
\end{tabular}

%% file: ms_3_Table2b.tex
\begin{tabular}{|l|c|c|c|c|c|c|c|c|}
\hline
& \multicolumn{2}{|c|}{} & \multicolumn{2}{|c|}{} &
\multicolumn{4}{|c|}{\bf MSML$^{\,\rm c}$}
\\
\cline{6-9}
${}^{{}^{\bf\displaystyle Variable}}$ &
\multicolumn{2}{|c|}{${}^{{}^{\bf\displaystyle ML\mbox{\bf-}by\mbox{\bf-}MLE^{\,\rm a}}}$} & \multicolumn{2}{|c|}{${}^{{}^{\bf\displaystyle ML\mbox{\bf-}by\mbox{\bf-}MCMC^{\,\rm b}}}$} &
\multicolumn{2}{|c|}{{\bf state }{\boldmath$s=0$}} &
\multicolumn{2}{|c|}{{\bf state }{\boldmath$s=1$}}
\\
\cline{2-9}
& {\bf fatality} & {\bf injury} & {\bf fatality} & {\bf injury}
& {\bf fatality} & {\bf injury} & {\bf fatality} & {\bf injury}
\\
\hline
Probability of severity outcome [$P_{t,n}^{(i)}$
given by Eq.~(\ref{EQ_ML})], averaged
& & & & & & & &\\
over all values of explanatory variables ${\bf X}_{t,n}$ & -- &
-- &
$.00747$ &
$.179$ &
$.00823$ &
$.183$ &
$.00218$ &
$.158$\\
\hline
Markov transition probability of jump $0\to1$ ($p_{0\rightarrow1}$) &
\multicolumn{2}{|c|}{--} &
\multicolumn{2}{|c|}{--} &
\multicolumn{4}{|c|}{$.0767^{.157}_{.0269}$}\\
\hline
Markov transition probability of jump $1\to0$ ($p_{1\rightarrow0}$) &
\multicolumn{2}{|c|}{--} &
\multicolumn{2}{|c|}{--} &
\multicolumn{4}{|c|}{$.613^{.864}_{.337}$}\\
\hline
Unconditional probabilities of states 0 and 1 (${\bar p}_0$ and ${\bar p}_1$) &
\multicolumn{2}{|c|}{--} &
\multicolumn{2}{|c|}{--} &
\multicolumn{4}{|c|}{$.887^{.959}_{.770}$\quad and\quad $.113^{.230}_{.0409}$}\\
\hline
\hline
Total number of free model parameters ($\beta$-s) &
\multicolumn{2}{|c|}{$24$} &
\multicolumn{2}{|c|}{$24$} &
\multicolumn{4}{|c|}{$25$}\\
\hline
Posterior average of the log-likelihood (LL) &
\multicolumn{2}{|c|}{--} &
\multicolumn{2}{|c|}{$-7406.39^{-7400.61}_{-7414.03}$} &
\multicolumn{4}{|c|}{$-7349.06^{-7335.46}_{-7364.47}$}\\
\hline
Max$(LL)$:\quad estimated max.~log-likelihood (LL) for MLE; &
\multicolumn{2}{|c|}{} & \multicolumn{2}{|c|}{} &
\multicolumn{4}{|c|}{} \\
maximum observed value of LL for Bayesian-MCMC & \multicolumn{2}{|c|}{$-7384.05\,{\rm(MLE)}\!$} &
\multicolumn{2}{|c|}{$-7396.37\,{\rm(observed)}\!$} &
\multicolumn{4}{|c|}{$-7318.21\,{\rm(observed)}\!$}\\
\hline
Logarithm of marginal likelihood of data ($\ln[f({\bf Y}|{\cal M})]$) &
\multicolumn{2}{|c|}{--} &
\multicolumn{2}{|c|}{$-7417.98^{-7413.72}_{-7420.23}$} &
\multicolumn{4}{|c|}{$-7377.49^{-7369.62}_{-7380.00}$}\\
\hline
Goodness-of-fit p-value & \multicolumn{2}{|c|}{--} &
\multicolumn{2}{|c|}{$0.337$} &
\multicolumn{4}{|c|}{$0.255$}\\
\hline
Maximum of the potential scale reduction factors (PSRF)$^{\,\rm d}$ &
\multicolumn{2}{|c|}{--} &
\multicolumn{2}{|c|}{$1.00319$} &
\multicolumn{4}{|c|}{$1.00073$}\\
\hline
Multivariate potential scale reduction factor (MPSRF)$^{\,\rm d}$ &
\multicolumn{2}{|c|}{--} &
\multicolumn{2}{|c|}{$1.00376$} &
\multicolumn{4}{|c|}{$1.00085$}\\
\hline
Number of available observations & \multicolumn{8}{|c|}
{accidents = fatalities + injuries + PDOs:\quad\quad\quad
$17797=138+3184+14485$}\\
\hline
\multicolumn{9}{l}{$^{\rm a}$ Standard (conventional)
multinomial logit (ML) model estimated by maximum likelihood
estimation (MLE).}\\
\multicolumn{9}{l}{$^{\rm b}$ Standard multinomial logit (ML)
model estimated by Markov Chain Monte Carlo (MCMC) simulations.}\\
\multicolumn{9}{l}{$^{\rm c}$ Two-state Markov switching
multinomial logit (MSML) model estimated by Markov Chain Monte Carlo
(MCMC) simulations.}\\
\multicolumn{9}{l}{$^{\rm d}$ PSRF/MPSRF are calculated separately/jointly
for all continuous model parameters. PSRF and MPSRF are close to 1 for
converged MCMC chains.}\\
\end{tabular}

%% file: ms_3_Table3a.tex
\tabcolsep=0.3em
\renewcommand{\arraystretch}{1.45}

\begin{tabular}{|l|c|c|c|c|c|c|c|c|}
\hline
& \multicolumn{2}{|c|}{} & \multicolumn{2}{|c|}{} &
\multicolumn{4}{|c|}{\bf MSML$^{\,\rm c}$}
\\
\cline{6-9}
${}^{{}^{\bf\displaystyle Variable}}$ &
\multicolumn{2}{|c|}{${}^{{}^{\bf\displaystyle ML\mbox{\bf-}by\mbox{\bf-}MLE^{\,\rm a}}}$} & \multicolumn{2}{|c|}{${}^{{}^{\bf\displaystyle ML\mbox{\bf-}by\mbox{\bf-}MCMC^{\,\rm b}}}$} &
\multicolumn{2}{|c|}{{\bf state }{\boldmath$s=0$}} &
\multicolumn{2}{|c|}{{\bf state }{\boldmath$s=1$}}
\\
\cline{2-9}
& {\bf fatality} & {\bf injury} & {\bf fatality} & {\bf injury}
& {\bf fatality} & {\bf injury} & {\bf fatality} & {\bf injury}
\\
\hline
Intercept (constant term) & $-3.98^{-3.66}_{-4.30}$ &
$-1.67^{-1.53}_{-1.80}$ &
$-4.03^{-3.71}_{-4.36}$ &
$-1.71^{-1.58}_{-1.85}$ &
$-3.44^{-3.10}_{-3.79}$ &
$-1.68^{-1.54}_{-1.81}$ &
$-4.96^{-4.15}_{-5.96}$ &
$-1.68^{-1.54}_{-1.81}$\\
\hline
Summer season (dummy) & $.232^{.307}_{.156}$ &
$.232^{.307}_{.156}$ &
$.232^{.307}_{.157}$ &
$.232^{.307}_{.157}$ &
$.238^{.314}_{.163}$ &
$.238^{.314}_{.163}$ &
$.238^{.314}_{.163}$ &
$.238^{.314}_{.163}$\\
\hline
Roadway type (dummy: 1 if urban, 0 if rural) &
$-.390^{-.302}_{-.478}$ &
$-.390^{-.302}_{-.478}$ &
$-.395^{-.306}_{-.483}$ &
$-.395^{-.306}_{-.483}$ &
-- &
$-.385^{-.296}_{-.474}$ &
$-2.05^{-.954}_{-3.62}$ &
$-3.85^{-.296}_{-.474}$\\
\hline
Daylight or street lights are lit up if dark (dummy) &
$-.646^{-.408}_{-.884}$ &
$.193^{.261}_{.125}$ &
$-.641^{-.404}_{-.879}$ &
$.199^{.267}_{.132}$ &
$-.689^{-.448}_{-.931}$ &
-- &
$-.689^{-.448}_{-.931}$ &
$.277^{.378}_{.177}$\\
\hline
Precipitation: rain/freezing rain/snow/sleet/hail (dummy) &
$-.854^{.466}_{-1.24}$ &
-- &
$-.868^{-.494}_{-1.27}$ &
-- &
$-.829^{-.448}_{-1.24}$ &
-- &
$-.829^{-.448}_{-1.24}$ &
--\\
\hline
Roadway median is drivable (dummy) &
$-.583^{-.225}_{-.940}$ &
-- &
$-.596^{-.250}_{-.964}$ &
-- &
$-.589^{-.241}_{-.960}$ &
-- &
$-.589^{-.241}_{-.960}$ &
--\\
\hline
Roadway is straight (dummy) &
$-.284^{-.214}_{-.353}$ &
$-.284^{-.214}_{-.353}$ &
$-.283^{-.214}_{-.352}$ &
$-.283^{-.214}_{-.352}$ &
$-.117^{-.0184}_{-.214}$ &
$-.117^{-.0184}_{-.214}$ &
$-.117^{-.0184}_{-.214}$ &
$-.465^{-.360}_{-.573}$\\
\hline
Primary cause of the accident is environment-related (dummy) &
$-4.23^{-3.59}_{-4.86}$ &
$-1.83^{-1.76}_{-1.91}$ &
$-4.28^{-3.67}_{-4.97}$ &
$-1.84^{-1.76}_{-1.91}$ &
$-4.40^{-3.79}_{-5.10}$ &
$-2.30^{-2.16}_{-2.44}$ &
$-4.40^{-3.79}_{-5.10}$ &
$-1.41^{-1.26}_{-1.55}$\\
\hline
Help arrived in 20 minutes or less after the crash (dummy) &
$.840^{.917}_{.762}$ &
$.840^{.917}_{.762}$ &
$.863^{.945}_{.781}$ &
$.863^{.945}_{.781}$ &
-- &
$.861^{.944}_{.778}$ &
$1.64^{2.64}_{.856}$ &
$.861^{.944}_{.778}$\\
\hline
The vehicle at fault is a motorcycle (dummy) &
$3.10^{3.31}_{2.89}$ &
$3.10^{3.31}_{2.89}$ &
$3.10^{3.31}_{2.89}$ &
$3.10^{3.31}_{2.89}$ &
$3.37^{3.66}_{3.09}$ &
$3.37^{3.66}_{3.09}$ &
$3.37^{3.66}_{3.09}$ &
$2.82^{3.19}_{2.47}$\\
\hline
Number of occupants in the vehicle at fault &
$.0557^{.0850}_{.0265}$ &
$.0557^{.0850}_{.0265}$ &
$.0565^{.0858}_{.0276}$ &
$.0565^{.0858}_{.0276}$ &
$.0942^{.138}_{.0528}$ &
$.0942^{.138}_{.0528}$ &
$.0942^{.138}_{.0528}$ &
--\\
\hline
At least one of the vehicles involved was on fire (dummy) &
$1.90^{2.45}_{1.33}$ &
$.456^{.780}_{.133}$ &
$1.87^{2.42}_{1.28}$ &
$.447^{.768}_{.124}$ &
$1.87^{2.43}_{1.28}$ &
$.461^{.782}_{.137}$ &
$1.87^{2.43}_{1.28}$ &
$.461^{.782}_{.137}$\\
\hline
Age of the driver at fault (in years) &
$\displaystyle{14.6^{21.4\vphantom{O^o}}_{7.80}}\atop\displaystyle{{}\times10^{-3}}$ &
$\displaystyle{-2.80^{-.800\vphantom{O^o}}_{-4.70}}\atop\displaystyle{{}\times10^{-3}}$ &
$\displaystyle{14.5^{21.3\vphantom{O^o}}_{7.67}}\atop\displaystyle{{}\times10^{-3}}$ &
$\displaystyle{-2.71^{-.723\vphantom{O^o}}_{-4.69}}\atop\displaystyle{{}\times10^{-3}}$ &
$\displaystyle{14.5^{21.4\vphantom{O^o}}_{7.63}}\atop\displaystyle{{}\times10^{-3}}$ &
$\displaystyle{-2.46^{-.469\vphantom{O^o}}_{-4.44}}\atop\displaystyle{{}\times10^{-3}}$ &
$\displaystyle{14.5^{21.4\vphantom{O^o}}_{7.63}}\atop\displaystyle{{}\times10^{-3}}$ &
$\displaystyle{-2.46^{-.469\vphantom{O^o}}_{-4.44}}\atop\displaystyle{{}\times10^{-3}}$\\
\hline
Gender of the driver at fault (dummy) &
$-.496^{-.211}_{-.780}$ &
$.279^{.344}_{.214}$ &
$-.505^{-.225}_{-.794}$ &
$.278^{.343}_{.213}$ &
$-.473^{-.192}_{-.764}$ &
$.283^{.348}_{.218}$ &
$-.473^{-.192}_{-.764}$ &
$.283^{.348}_{.218}$\\
\hline
Age of the vehicle at fault (in years) & -- &
$.0334^{.0392}_{.0276}$ &
-- &
$.0335^{.0393}_{.0277}$ &
-- &
$.0332^{.0390}_{.0274}$ &
-- &
$.0332^{.0390}_{.0274}$\\
\hline
license state of the vehicle at fault is a U.S. state
except Indiana & & & & & & & &\\
and its neighboring states (IL, KY, OH, MI)" indicator variable & -- &
$-.449^{-.217}_{-.681}$ &
-- &
$-.444^{-.217}_{-.679}$ &
-- &
$-.436^{-.208}_{-.671}$ &
-- &
$-.436^{-.208}_{-.671}$\\
\hline
\end{tabular}

%% file: ms_3_Table3b.tex
\renewcommand{\arraystretch}{1.45}

\begin{tabular}{|l|c|c|c|c|c|c|c|c|}
\hline
& \multicolumn{2}{|c|}{} & \multicolumn{2}{|c|}{} &
\multicolumn{4}{|c|}{\bf MSML$^{\,\rm c}$}
\\
\cline{6-9}
${}^{{}^{\bf\displaystyle Variable}}$ &
\multicolumn{2}{|c|}{${}^{{}^{\bf\displaystyle ML\mbox{\bf-}by\mbox{\bf-}MLE^{\,\rm a}}}$} & \multicolumn{2}{|c|}{${}^{{}^{\bf\displaystyle ML\mbox{\bf-}by\mbox{\bf-}MCMC^{\,\rm b}}}$} &
\multicolumn{2}{|c|}{{\bf state }{\boldmath$s=0$}} &
\multicolumn{2}{|c|}{{\bf state }{\boldmath$s=1$}}
\\
\cline{2-9}
& {\bf fatality} & {\bf injury} & {\bf fatality} & {\bf injury}
& {\bf fatality} & {\bf injury} & {\bf fatality} & {\bf injury}
\\
\hline
Probability of severity outcome [$P_{t,n}^{(i)}$
given by Eq.~(\ref{EQ_ML})], averaged
& & & & & & & &\\
over all values of explanatory variables ${\bf X}_{t,n}$ & -- &
-- &
$.0089$ &
$.179$ &
$.00951$ &
$.180$ &
$.00804$ &
$.179$\\
\hline
Markov transition probability of jump $0\to1$ ($p_{0\rightarrow1}$) &
\multicolumn{2}{|c|}{--} &
\multicolumn{2}{|c|}{--} &
\multicolumn{4}{|c|}{$.335^{.465}_{.216}$}\\
\hline
Markov transition probability of jump $1\to0$ ($p_{1\rightarrow0}$) &
\multicolumn{2}{|c|}{--} &
\multicolumn{2}{|c|}{--} &
\multicolumn{4}{|c|}{$.450^{.610}_{.313}$}\\
\hline
Unconditional probabilities of states 0 and 1 (${\bar p}_0$ and ${\bar p}_1$) &
\multicolumn{2}{|c|}{--} &
\multicolumn{2}{|c|}{--} &
\multicolumn{4}{|c|}{$.574^{.681}_{.504}$\quad and\quad $.426^{.496}_{.319}$}\\
\hline
\hline
Total number of free model parameters ($\beta$-s) &
\multicolumn{2}{|c|}{$22$} &
\multicolumn{2}{|c|}{$22$} &
\multicolumn{4}{|c|}{$28$}\\
\hline
Posterior average of the log-likelihood (LL) &
\multicolumn{2}{|c|}{--} &
\multicolumn{2}{|c|}{$-13867.40^{-13861.92}_{-13874.73}$} &
\multicolumn{4}{|c|}{$-13781.76^{-13765.02}_{-13800.89}$}\\
\hline
Max$(LL)$:\quad estimated max.~log-likelihood (LL) for MLE; &
\multicolumn{2}{|c|}{} & \multicolumn{2}{|c|}{} &
\multicolumn{4}{|c|}{} \\
maximum observed value of LL for Bayesian-MCMC &
\multicolumn{2}{|c|}{$-13846.60\,{\rm(MLE)}\!$} &
\multicolumn{2}{|c|}{$-13858.00\,{\rm(observed)}\!$} &
\multicolumn{4}{|c|}{$-13745.61\,{\rm(observed)}\!$}\\
\hline
Logarithm of marginal likelihood of data ($\ln[f({\bf Y}|{\cal M})]$) &
\multicolumn{2}{|c|}{--} &
\multicolumn{2}{|c|}{$-13877.89^{-13874.24}_{-13880.38}$} &
\multicolumn{4}{|c|}{$-13820.20^{-13808.85}_{-13821.73}$}\\
\hline
Goodness-of-fit p-value & \multicolumn{2}{|c|}{--} &
\multicolumn{2}{|c|}{$0.515$} &
\multicolumn{4}{|c|}{$0.445$}\\
\hline
Maximum of the potential scale reduction factors (PSRF)$^{\,\rm d}$ &
\multicolumn{2}{|c|}{--} &
\multicolumn{2}{|c|}{$1.00027$} &
\multicolumn{4}{|c|}{$1.00029$}\\
\hline
Multivariate potential scale reduction factor (MPSRF)$^{\,\rm d}$ &
\multicolumn{2}{|c|}{--} &
\multicolumn{2}{|c|}{$1.00041$} &
\multicolumn{4}{|c|}{$1.00045$}\\
\hline
Number of available observations & \multicolumn{8}{|c|}
{accidents = fatalities + injuries + PDOs:\quad\quad\quad$33528=302+6018+27208$}\\
\hline
\multicolumn{9}{l}{$^{\rm a}$ Standard (conventional)
multinomial logit (ML) model estimated by maximum likelihood
estimation (MLE).}\\
\multicolumn{9}{l}{$^{\rm b}$ Standard multinomial logit (ML)
model estimated by Markov Chain Monte Carlo (MCMC) simulations.}\\
\multicolumn{9}{l}{$^{\rm c}$ Two-state Markov switching
multinomial logit (MSML) model estimated by Markov Chain Monte Carlo
(MCMC) simulations.}\\
\multicolumn{9}{l}{$^{\rm d}$ PSRF/MPSRF are calculated separately/jointly
for all continuous model parameters. PSRF and MPSRF are close to 1 for
converged MCMC chains.}\\
\end{tabular}

%% file: ms_3_Table4a.tex
\tabcolsep=0.5em
\begin{tabular}{|l|l|}
\hline
\bf Variable & \hspace{4cm}{\bf Description}\\
\hline
$\langle P_{t,n}^{(i)}\rangle_X$ & probability of $i^{\rm th}$ severity
outcome averaged over all values of explanatory variables ${\bf X}_{t,n}$\\
\hline
$p_{0\rightarrow1}$ & Markov transition probability of jump from
state 0 to state 1 as time $t$ increases to $t+1$\\
\hline
$p_{1\rightarrow0}$ & Markov transition probability of jump from
state 1 to state 0 as time $t$ increases to $t+1$\\
\hline
${\bar p}_0$ and ${\bar p}_1$ & unconditional probabilities of
states 0 and 1\\
\hline
$\#$ free par. & total number of free model coefficients ($\beta$-s)\\
\hline
averaged $LL\!$ & posterior average of the log-likelihood (LL)\\
\hline
$\max(LL)$ & for MLE it is the maximal value of $LL$ at convergence;
for Bayesian-MCMC estimation \\
& it is the maximal observed value of LL during the MCMC simulations\\
\hline
marginal $LL\!$ & logarithm of marginal likelihood of data,
$\ln[f({\bf Y}|{\cal M})]$, given model ${\cal M}$\\
\hline
max(PSRF) & maximum of the potential scale reduction factors (PSRF)
calculated separately for all \\
& continuous model parameters, PSRF is close to 1 for converged
MCMC chains\\
\hline
MPSRF & multivariate PSRF calculated jointly for all parameters,
close to 1 for converged MCMC\\
\hline
accept. rate & average rate of acceptance of candidate values
during Metropolis-Hasting MCMC draws\\
\hline
$\#$ observ. & number of observations of accident severity
outcomes available in the data sample\\
\hline
\hline
age0 & "age of the driver at fault is $<18$ years" indicator variable
(dummy)\\
\hline
age0o & "age of the oldest driver involved into the accident is $<18$ years"
indicator variable\\
\hline
cons & "construction at the accident location" indicator variable\\
\hline
curve & "road is at curve" indicator variable\\
\hline
dark & "dark time with no street lights" indicator variable\\
\hline
darklamp & "dark AND street lights on" indicator variable\\
\hline
day & "daylight" indicator variable\\
\hline
dayt & "day hours: 9:00 to 17:00" indicator variable\\
\hline
driv & "road median is drivable" indicator variable\\
\hline
driver & "primary cause of the accident is driver-related" indicator
variable\\
\hline
dry & "roadway surface is dry" indicator variable\\
\hline
env & "primary cause of the accident is environment-related" indicator
variable\\
\hline
fog & "fog OR smoke OR smog" indicator variable\\
\hline
hl10 & "help arrived in 10 minutes or less after the crash" indicator
variable\\
\hline
hl20 & "help arrived in 20 minutes or less after the crash" indicator
variable\\
\hline
Ind & "license state of the vehicle at fault is Indiana" indicator
variable\\
\hline
intercept & "constant term (intercept)" quantitative variable\\
\hline
jobend & "after work hours: from 16:00 to 19:00" indicator variable\\
\hline
light & "daylight OR street lights are lit up if dark"
indicator variable\\
\hline
maxpass & "the largest number of occupants in all vehicles involved"
quantitative variable\\
\hline
mm & "two male drivers are involved" indicator
variable (used only if a 2-vehicle accident)\\
\hline
morn & "morning hours: 5:00 to 9:00" indicator variable\\
\hline
moto & "the vehicle at fault is a motorcycle" indicator variable\\
\hline
\end{tabular}

%% file: ms_3_Table4b.tex
\tabcolsep=0.5em
\begin{tabular}{|l|l|}
\hline
\bf Variable & \hspace{4cm}{\bf Description}\\
\hline
nigh & "late night hours: 1:00 to 5:00" indicator variable\\
\hline
nocons & "no construction at the accident location" indicator variable\\
\hline
nojun & "no road junction at the accident location" indicator variable\\
\hline
nonroad & "non-roadway crash (parking lot, etc.)" indicator variable\\
\hline
nosig & "no any traffic control device for the vehicle at fault"
indicator variable\\
\hline
olddrv & "the driver at fault is older than the other driver"
indicator var. (if a 2-vehicle accident)\\
\hline
oldvage & "age (in years) of the oldest vehicle involved" indicator variable\\
\hline
othUS & "license state of the vehicle at fault is a U.S. state except
Indiana and its neighboring\\
& states (IL, KY, OH, MI)" indicator variable\\
\hline
precip & "precipitation: rain OR snow OR sleet OR hail OR freezing rain"
indicator variable\\
\hline
priv & "road traveled by the vehicle at fault is a private drive"
indicator variable\\
\hline
r21 & "road traveled by the vehicle at fault is two-lane AND one-way"
indicator variable\\
\hline
rmd2 & "road traveled by the vehicle at fault is multi-lane AND divided
two-way" indicator var.\\
\hline
singSUV & "one of the two vehicles involved is a pickup OR a van OR 
a sport utility vehicle"\\
& indicator variable (used only if a 2-vehicle accident)\\
\hline
singTR & "one of the two vehicles is a truck OR a tractor"
indicator var. (if a 2-vehicle accident)\\
\hline
slush & "roadway surface is covered by snow/slush" indicator variable\\
\hline
snow & "snowing weather" indicator variable\\
\hline
str & "road is straight" indicator variable\\
\hline
sum & "summer season" indicator variable\\
\hline
sund & "Sunday" indicator variable\\
\hline
thday & "Thursday" indicator variable\\
\hline
vage & "age (in years) of the vehicle at fault" quantitative variable\\
\hline
veh & "primary cause of accident is vehicle-related" indicator variable\\
\hline
voldg & "the vehicle at fault is more than 7 years old" indicator variable\\
\hline
voldo & "age of the oldest vehicle involved is more than 7 years" indicator
variable\\
\hline
wall & "road median is a wall" indicator variable\\
\hline
way4 & "accident location is at a 4-way intersection" indicator variable\\
\hline
wint & "winter season" indicator variable\\
\hline
$X_{12}$ & "road type" indicator variable (1 if urban, 0 if rural)\\
\hline
$X_{27}$ & "number of occupants in the vehicle at fault"
quantitative variable\\
\hline
$X_{29}$ & "speed limit" quantitative var. (used if known and
the same for all vehicles involved)\\
\hline
$X_{33}$ & "at least one of the vehicles involved was on fire"
indicator variable\\	
\hline
$X_{34}$ & "age (in years) of the driver at fault" quantitative
variable\\
\hline
$X_{35}$ & "gender of the driver at fault" indicator
variable (1 if female, 0 if male)\\
\hline
\end{tabular}

%% file: ms_3_Table5.tex
\tabcolsep=0.5em
\begin{tabular}{|l|c|c|c|c|c|c|}
\hline
& 1-vehicle,  & 1-vehicle, & 1-vehicle,   & 1-vehicle,   & 1-vehicle, & 2-vehicle,\\
& interstates & US routes  & state routes & county roads & streets    & streets  \\
\hline
1-vehicle, interstates & $1$ & $0.418$ & $0.293$ & $0.606$ & $-0.013$ & $-0.173$\\
\hline
1-vehicle, US routes & $0.418$ & $1$ & $0.263$ & $0.688$ & $-0.070$ & $-0.155$\\
\hline
1-vehicle, state routes & $0.293$ & $0.263$ & $1$ & $0.409$ & $-0.047$ & $-0.035$\\
\hline
1-vehicle, county roads & $0.606$ & $0.688$ & $0.409$ & $1$ & $-0.022$ & $-0.051$\\
\hline
1-vehicle, streets & $-0.013$ & $-0.070$ & $-0.047$ & $-0.022$ & $1$ & $0.115$\\
\hline
2-vehicle, streets & $-0.173$ & $-0.155$ & $-0.035$ & $-0.051$ & $0.115$ & $1$\\
\hline
\hline
\multicolumn{7}{|c|}{All year}\\
\hline
Precipitation (inch) & $-0.139$ & $-0.060$ & $0.096$ & $-0.037$ & $0.067$ & $0.146$\\
\hline
Temperature ($^oF$) & $-0.606$ & $-0.439$ & $-0.234$ & $-0.665$ & $0.231$ & $0.220$\\
\hline
Snowfall (inch) & $0.479$ & $0.635$ & $0.319$ & $0.723$ & $0.003$ & $-0.100$\\
\cline{2-7}
$\quad\quad\quad>0.0$ (dummy) & $0.695$ & $0.412$ & $0.382$ & $0.695$ & $-0.142$ & $-0.131$\\
\cline{2-7}
$\quad\quad\quad>0.1$ (dummy) & $0.532$ & $0.585$ & $0.328$ & $0.847$ & $-0.046$ & $-0.161$\\
\hline
Wind gust (mph) & $0.108$ & $0.100$ & $0.087$ & $0.206$ & $0.164$ & $0.051$\\
\hline
Fog / Frost (dummy) & $0.093$ & $0.164$ & $0.193$ & $0.167$ & $0.047$ & $0.119$\\
\hline
Visibility distance (mile) & $-0.228$ & $-0.221$ & $-0.172$ & $-0.298$ & $-0.019$ & $-0.081$\\
\hline
\hline
\multicolumn{7}{|c|}{Winter (November - March)}\\
\hline
Precipitation (inch) & $-0.134$ & $-0.037$ & $0.027$ & $-0.053$ & $0.065$ & $0.356$\\
\hline
Temperature ($^oF$) & $-0.595$ & $-0.479$ & $-0.397$ & $-0.735$ & $-0.008$ & $0.236$\\
\hline
Snowfall (inch) & $0.439$ & $0.592$ & $0.375$ & $0.645$ & $0.157$ & $-0.110$\\
\cline{2-7}
$\quad\quad\quad>0.0$ (dummy) & $0.596$ & $0.282$ & $0.475$ & $0.607$ & $0.115$ & $-0.142$\\
\cline{2-7}
$\quad\quad\quad>0.1$ (dummy) & $0.445$ & $0.518$ & $0.370$ & $0.789$ & $0.112$ & $-0.210$\\
\hline
Wind gust (mph) & $0.302$ & $0.134$ & $0.122$ & $0.353$ & $0.237$ & $0.071$\\
\hline
Frost (dummy) & $0.537$ & $0.544$ & $0.440$ & $0.716$ & $0.052$ & $-0.225$\\
\hline
Visibility distance (mile) & $-0.251$ & $-.304$ & $-0.249$ & $-0.380$ & $-0.155$ & $-0.109$\\
\hline
\hline
\multicolumn{7}{|c|}{Summer (May - September)}\\
\hline
Precipitation (inch) & $0.000$ & $0.006$ & $0.259$ & $0.096$ & $0.047$ & $-0.063$\\
\hline
Temperature ($^oF$) & $0.179$ & $0.149$ & $0.113$ & $0.037$ & $0.062$ & $0.155$\\
\hline
Snowfall (inch) & -- & -- & -- & -- & -- & --\\
\cline{2-7}
$\quad\quad\quad>0.0$ (dummy) & -- & -- & -- & -- & -- & --\\
\cline{2-7}
$\quad\quad\quad>0.1$ (dummy) & -- & -- & -- & -- & -- & --\\
\hline
Wind gust (mph) & $-0.126$ & $-.009$ & $0.164$ & $0.029$ & $0.121$ & $0.034$\\
\hline
Fog (dummy) & $0.203$ & $0.193$ & $0.275$ & $0.101$ & $-0.076$ & $-0.011$\\
\hline
Visibility distance (mile) & $-0.139$ & $-0.124$ & $-0.062$ & $-0.009$ & $0.077$ & $-0.094$\\
\hline
\end{tabular}